# A Fiber Optic Based High Voltage System for Stellar Intensity Interferometry Observations

**Rylee Cardon, Nolan Matthews\*, A. Udara Abeysekara and David Kieda**
*Department of Physics and Astronomy*
*University of Utah*
*Salt Lake City, UT 84112, United States of America*
E-mail: rylee.cardon@gmail.com, nolankmatthews@gmail.com, udarabeysekara@yahoo.com, dave.kieda@utah.edu

Beginning in Fall 2018, the VERITAS high energy gamma-ray observatory (Amado, AZ) was upgraded to enable Stellar Intensity Interferometry (SII) observations during bright moon conditions. The system potentially allows VERITAS to spatially characterize stellar objects at visible wavelengths with sub-milliarcsecond angular resolution. This research project was on the construction of a high voltage power supply for the photomultiplier tubes (PMTs) used in the SII camera. The high voltage supply was designed to be electrically isolated from all other electronics (except for the PMT) to reduce noise pickup. The HV supply operates on a Li-Ion battery, and the high voltage level is remotely programmed using a pulse width modulation (PWM) signal that is generated by an Arduino Yun microcontroller and distributed through a fiber optic cable. The electrical isolation of the fiber optic control system suppresses the pickup of radio frequency interference through ground loops. A separate fiber optic transceiver pair is used for the on-off control of the high voltage power supply. Tests were performed that show the high voltage level is reproducible to within one volt for a given duty cycle of the PWM signal. Furthermore, the high voltage output level was shown to be stable with respect to variations in the input battery voltage used to power the high voltage supply. The high voltage system is currently being used in regular SII observations at VERITAS. This poster will describe the detailed design and performance of the system.

---

\* Speaker

**Introduction**

A Stellar Intensity Interferometry (SII) system has been implemented on the four VERITAS telescopes [1] to perform high angular resolution measurements of stars. Each telescope has been outfitted with a removable SII focal plane plate that utilizes a single photomultiplier tube [2] (PMT) that detect the light intensity fluctuations of stars at nanosecond timescales. The signals from each PMT are continuously streamed to disk, and then after observations are over, the recorded signals from each telescope are cross-correlated. The results from the correlated data provide a way to spatially characterize stellar sources (e.g., the general size and shape of the stars). The quality of the SII measurements depends upon the fidelity of the cross-correlations. Radio-frequency (RF) interference pickup in the PMT signals that can create unwanted spurious correlations that will degrade the sensitivity of the SII instrument.

A low-noise high voltage (HV) power supply is essential for the reduction of noise pickup in the PMTs. Operation. The HV system must also allow remote control HV level that is delivered to the PMT so that the gain of the PMT can be adjusted. The remote operation allows for an adjustable detector gain that is needed to observe over a wide range of stellar magnitudes/fluxes. For bright stars, the PMT gain is set lower to avoid a significantly high anode current that would degrade the PMT response over time, whereas, for dimmer stars, the gain is set higher to expand the PMT signal's dynamic range to match the range of the data acquisition digitizers. The HV system was designed to minimize the effect of RF pickup by incorporating fiber-optic cables to provide electrical isolation, and Li-Ion battery power to eliminate RF pickup in ground loops. The system was tested in the laboratory to demonstrate its successful operation over nightly time-scales and is now in place on the VERITAS telescopes.

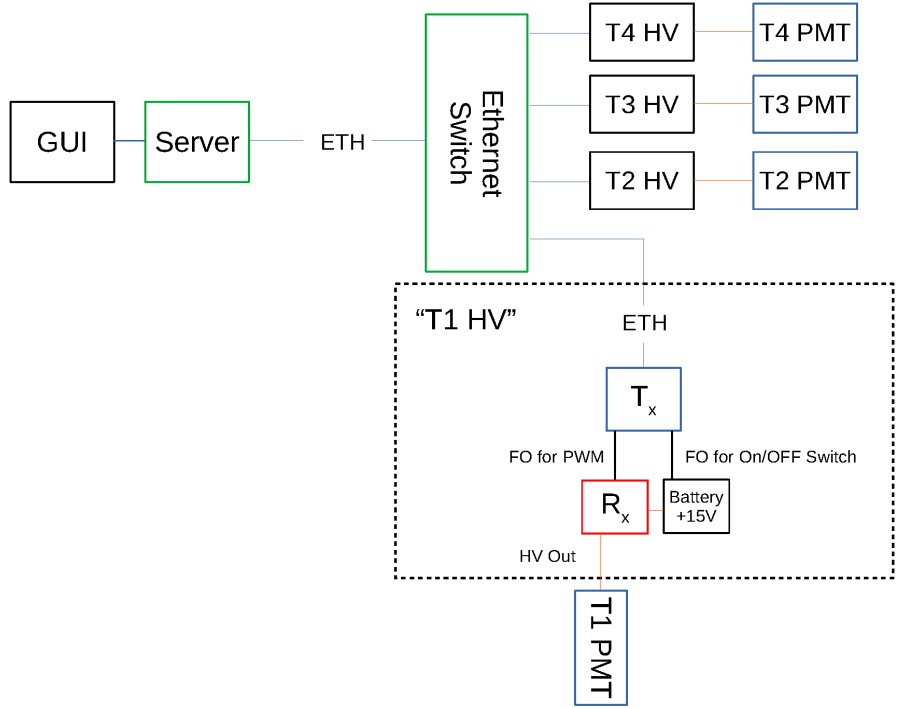

*Figure 1:* Block diagram of the SII-HV system. The HV control GUI is hosted on a server computer which communicates to the four high voltage systems (T1, T2, T3, T4) via Ethernet. Inside the HV system at each telescope (e.g., T1 HV), server commands are received by an Arduino Yun microcontroller ($T_x$) which generates a PWM signal and On/Off signal which is transmitted to the HV control system ($R_x$) and battery at the camera focal plane. The HV control system $R_x$ supplied the HV to the Photomultiplier tube (PMT).

## 2. Design of the SII-HV System

A block diagram showing the overall design of the SII-HV system is presented in Figure 3. A PHP-based GUI allows the user to independently set the amount of HV that is delivered to each PMT. The GUI is hosted as a web page on a local computer (Figure 2) which receives HTTP-based control commands from the GUI. The user-GUI generates HTTP strings that encode the desired values of the HV at each telescope and sends the commands to an Arduino Yun Microcontroller located in each Telescope Control room. The Yun uses the received HV values to control an electronic pulse-width-modulation (PWM) signal output on the Yun card. The output of this PWM signal drives an FB129-ND fiber-optic transmitter to generate a PWM optical signal. The optical PWM signal is sent over a long plastic fiber-optic cable to each telescope (Figure 2). The control system uses inexpensive Eska GHCP4002 Premier 2.2mm core plastic fiber. At the focal plane of each camera, a custom-built HV supply receives the PWM optical signal from the fiber optic cable and converts the signal to a DC voltage level (0-5V) using a low pass filter (Figure 3). This DC control voltage sets the HV output level of a DC-to-DC HV converter module (EMCO CA-12N). The output of this module directly drives the PMT bleeder chain. A high capacity Li-Ion battery is mounted separately in the camera and provides +15 V to power the receiver unit. During the daytime, a Li-Ion battery charger, plugged into a 120V AC socket in each camera, recharges the HV battery to full power.

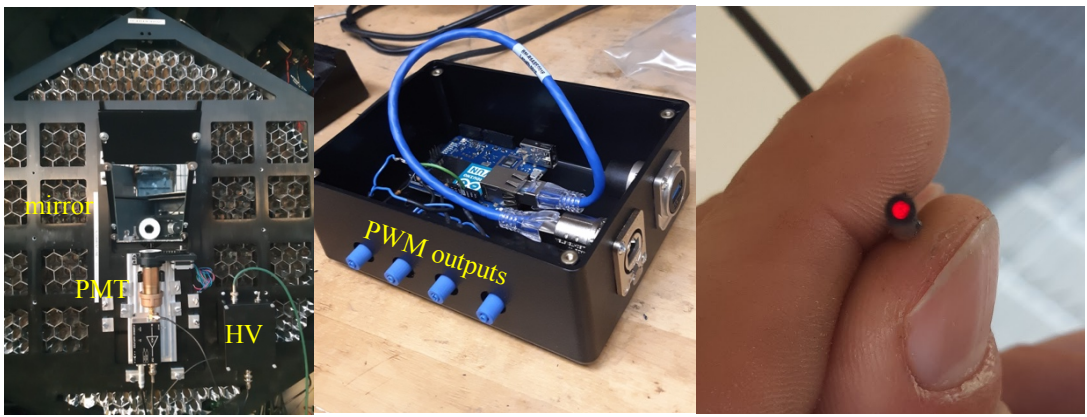

***Figure 2: Left:*** *PMT setup at camera focal plane with 45º mirror and remotely controlled HV supply (bottom left corner).* ***Middle:*** *PWM HV control transmitter box, including Arduino Yun, two fiber optic on-off controls, and two PWM outputs in each control box (blue connectors).* ***Right:*** *The ESKA GHCP 4002 2.2mm core plastic fiber optic cable that transmits the PWM signal from the telescope trailer to the HV supply at the camera Focal Plane plate.*

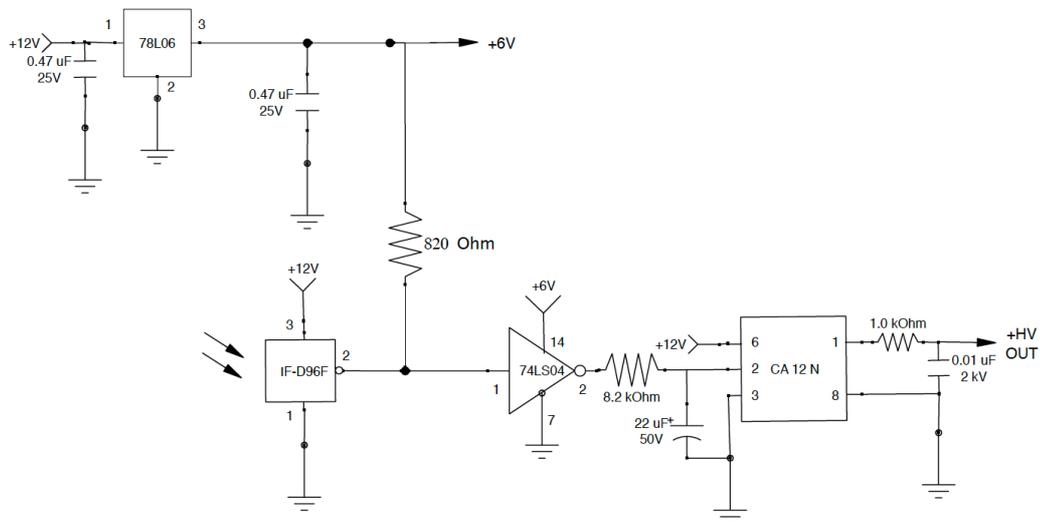

*Figure 3:* Circuit diagram for the SII-HV receiver located in the camera of the telescope. The low pass filter for converting the PWM to the DC signal is comprised of the 8.2 kOhm resistor and the 22 uF capacitor output of the LS04 inverter and feeding into pin 2 of the CA-12N DC-to-DC HV converter.

### 3. Laboratory Tests and Results

The stability of the SII-HV system was tested in the laboratory tests to determine its usefulness for the SII application. It is critical that the HV power supply output remains stable throughout the night, even when the Li-Ion battery is slowly being drained. An extended test was performed to measure the variation of the output of the SII-HV supply with respect to changes in the supply voltage of the 15V battery supply. This test was performed over several hours to quantify the long-term stability of the HV output. The SII-HV was set up to deliver a constant output HV into a simulated PMT load, and measurements of the input battery voltage and output HV level were made over regular intervals over several hours. Figure 4 shows the result of the laboratory tests. During a four hour test, the battery voltage drops at a linear rate of approximately 25 mV/hour. During the same interval, the HV output was seen to vary by less than 0.015% over the same period, or less than 0.0038% per hour. Because the PMT used for this application (Hamamatsu super-bialkalai R10560) uses eight dynode stages[2], the gain of the PMT varies as $V^7$. A fractional change in voltage dV/V will then cause a fractional change in PMT gain 7*(dV/V). We therefore expect the PMT gain will vary by less than ~0.01% over 4 hours, or less than 0.03% during an entire night's observation (12 hours).

By comparison, fluctuations in sky brightness can also affect PMT gains due to changing current level being drawn through the PMT bleeder chain. The Hamamatsu bleeder chain was designed to reduce the sensitivity of the PMT to changes in night sky brightness. The PMT temperature also has a small effect on the PMT gain. These effects are typically on the order of a fraction of a percent difference in gain under different operating conditions. Consequently, the variation in the PMT gain due to the Li-Ion battery drain is small compared to these other effects.

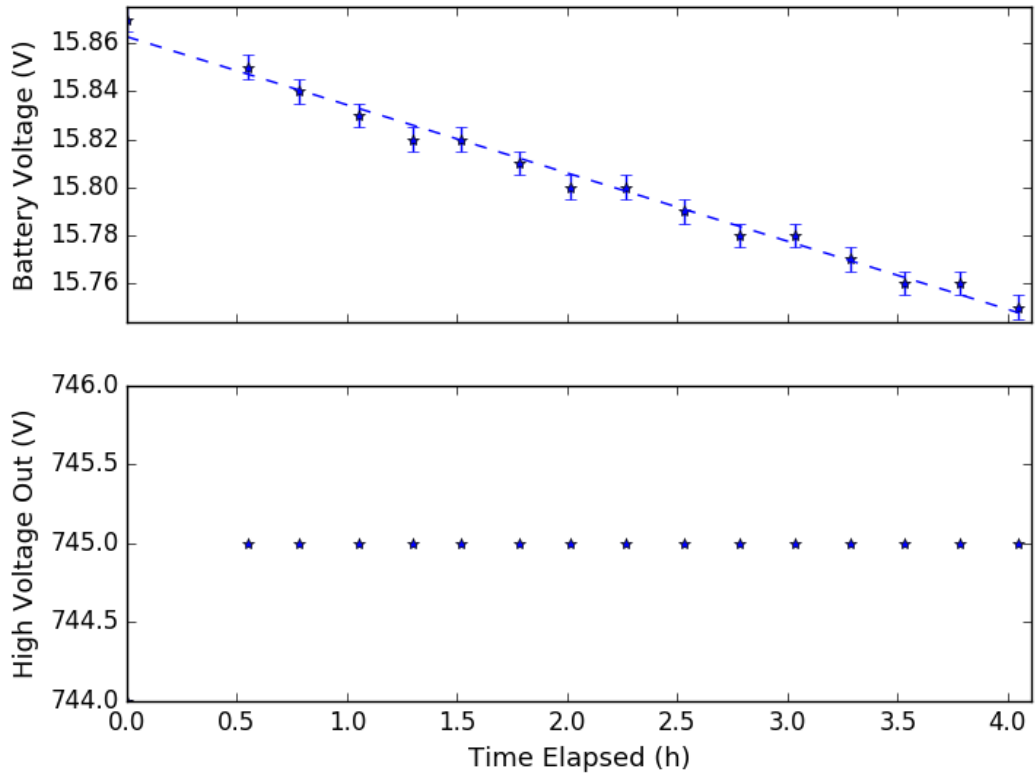

***Figure 4:*** *The above plot shows the results of tests on the SII-HV supply system demonstrating the stability of the system over time. The top panel shows the loss in battery voltage over time for the supply used to power the SII-HV receiver. A linear trend is fit to the data and shown by the dotted line. The bottom panel displays the measured HV level output from the SII-HV receiver as a function of time. Throughout the 4 hours, the output HV did not fluctuate at a level that was observable by the resolution of the measurement apparatus (< 0.1 V).*

### 4. Conclusions and Discussion

In these proceedings, a stable low-noise HV supply system for the SII system mounted onto the VERITAS telescopes are presented. The system was tested in the laboratory, and a HV stability of better than 0.015% (dV/V) was demonstrated over a four-hour interval. His corresponds to a relative gain stability dG/G of better than 0.01% The change in the PMT gain induced by the multi-hour battery drain is substantially smaller than gain shifts due to other factors associated with the PMT (sky brightness, temperature). Four remotely controllable HV systems are currently being used for on-sky SII observing at VERITAS during bright moon periods [5]. The overall gain and operation are stable over nightly timescales, thus demonstrating the working condition of the system. The HV system described here provides a low-cost, low-noise system use with PMT detectors for SII and for other low-noise applications.

We expect to implement further improvements to the SII-HV system in the future. The individual GUI system for each HV supply will be superseded by a single user interface with integrated data acquisition and high voltage control. Additionally, the electronic components in the receiver will be mounted onto a custom printed-circuit-board to improve the robustness and stability of the system. This improvement also has the benefit of making each module easily replaceable in the case of any failure in the system.

**Acknowledgments**

This research is supported by grants from the U.S. National Science Foundation. The authors gratefully acknowledge support under NSF Grant #AST 1806262 for the fabrication and commissioning of the VERITAS-SII instrumentation.